# SEGMENTATION OF TEMPOROMANDIBULAR JOINT STRUCTURES ON MRI IMAGES USING NEURAL NETWORKS FOR DIAGNOSIS OF PATHOLOGIES

Glukhikh[1] I. N., Karyakin[1] Y.E., Ivanov[1] M. I., Mendybaeva[1] O. E., Lebedev[2] A.V.
[1] University of Tyumen, [2] Tyumen center of Neurostomatology

**Abstract**

This article explores the use of artificial intelligence for the diagnosis of pathologies of the temporomandibular joint (TMJ), in particular, for the segmentation of the articular disc on MRI images. The relevance of the work is due to the high prevalence of TMJ pathologies, as well as the need to improve the accuracy and speed of diagnosis in medical institutions. During the study, the existing solutions (Diagnocat, MandSeg) were analyzed, which, as a result, are not suitable for studying the articular disc due to the orientation towards bone structures. To solve the problem, an original dataset was collected from 94 images with the classes "temporomandibular joint" and "jaw". To increase the amount of data, augmentation methods were used. After that, the models of U-Net, YOLOv8n, YOLOv11n and Roboflow neural networks were trained and compared. The evaluation was carried out according to the Dice Score, Precision, Sensitivity, Specificity, and Mean Average Precision metrics. The results confirm the potential of using the Roboflow model for segmentation of the temporomandibular joint. In the future, it is planned to develop an algorithm for measuring the distance between the jaws and determining the position of the articular disc, which will improve the diagnosis of TMJ pathologies**.**

## 1. Introduction

In recent years, the use of artificial intelligence (AI) has become an integral part of modern medicine. Automation of the process of diagnosis and treatment of diseases using machine learning algorithms can significantly improve the quality of medical care, speed up the process of processing large amounts of data and reduce the risk of errors associated with the human factor, such as doctor fatigue or subjectivity of opinions [1].

Modern diagnostic technologies based on AI are able not only to effectively evaluate and interpret medical images, including X-rays, MRI, CT and biometric data, but also to identify complex patterns that are not available for traditional methods of analysis [2, 3]. For example, neural network algorithms demonstrate high accuracy in detecting the early stages of cancer, which significantly increases the chances of patients for the timely initiation of therapy and a favorable prognosis [4, 5].

As is generally known, an artificial neural network is a conceptual structure for the development of AI algorithms. This is a model of the human brain, consisting of an interconnected network of neurons. Neurons are interconnected by communication channels and are able to quickly process incoming information and medical data [6].

One of the key and rapidly developing areas in medical practice, where the use of artificial intelligence is especially relevant and promising, is modern dentistry [7, 8]. The pathology of the temporomandibular joint continues to be a significant issue and subject of investigation in this field [9].

The TMJ is formed by the head of the lower jaw, crowning the condyle process, the mandibular fossa, the articular tubercle of the temporal bone, the joint capsule and the key element — the articular disc. It divides the articular cavity into two parts, which allows the jaw to perform complex movements: not only opening and closing the mouth, but also lateral movements, as well as forward and backward movements [10-12].

The main functions of the articular disc are to dampen movements in the joint, distribute the load on the bone structures and ensure smoothness during movement of the lower jaw. Due to its elasticity and strength, the disc minimizes friction between the bone surfaces, preventing their premature wear. Damage or dislocation of the articular disc can lead to a number of symptoms that cause discomfort to a person: clicks or crunches, pain, and limited jaw mobility [13-15].

According to numerous clinical observations and scientific publications, TMJ pathologies occur in 28-76% of patients seeking dental care in Russia. Despite the long period of studying diseases affecting the temporomandibular joint, this problem still remains relevant and requires more attention from doctors [16].

In modern medical institutions, it is not always possible to carry out detailed analyses of each image due to the high workload. In addition, not every doctor will be able to quickly and accurately diagnose an image. Early diagnosis of such disorders using machine learning algorithms is important, as it helps to prevent their development and reduce the risk of medical errors.

Currently, foreign researchers are just beginning to create solutions for studying the temporomandibular joint. Among the most progressive commercial products, the innovative Diagnocat service (Fig.1), developed by the company of the same name, should be highlighted [17]. This platform offers a fundamentally new approach to the analysis of maxillofacial structures. It includes fully automatic creation of 3D STL models and simultaneous segmentation of various anatomical structures such as soft tissues, teeth, maxillary sinus, mandibular canal, incisor canal, lower jaw, upper jaw, base of skull and respiratory tract. However, the key element of the TMJ, the articular disc, is a cartilaginous structure, which is why it cannot be studied using this service.

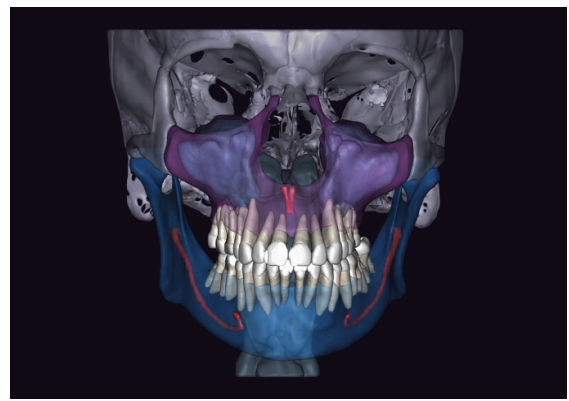

Figure 1 – Diagnocat service [17]

Another solution is MangSeg algorithm (Fig. 2), which combines image processing and machine learning approaches for automatic segmentation of condyles and branches of the mandible [18].

This solution focuses on the study of bone pathologies such as osteoarthritis, which means that it is also not suitable for studying the articular disc.

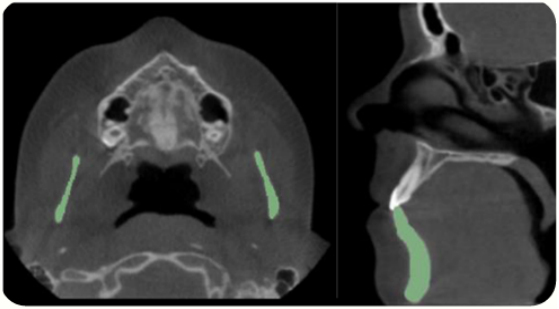

Figure 2 – Segmentation of mandibular branches using the MandSeg algorithm [18]

As you can see, most of the research is carried out on CT (computed tomography) and 3D images, however, we will work with MRI images, since they better show the cartilaginous structures, which is the TMJ disk.

The purpose of this study is to conduct a comparative analysis of the effectiveness of modern neural networks in the task of automated recognition of the temporomandibular joint on medical MRI images.

2. Research methods

2.1. Data collection and preparation

In the course of solving the problem, other studies on similar topics were found, but the datasets presented in them were either not publicly available or not suitable for the task. These were mainly CT scans, which did not show the key element of interest to us - the IVD. MRI scans of primates were also found, but their IVC structure is different from that of humans and is irrelevant to the task, so it was decided to form an original dataset.

The initial data was provided by the Center for Neurostomatology of Tyumen, it included 60 MRI scans in DICOM format. After a preliminary analysis, 94 axial slices were selected, on which the TMJ disc was clearly visualized. The conversion of images from DICOM to JPEG format was performed using the Radiant Dicom Viewer software. An example of an input image is shown in Figure 3.

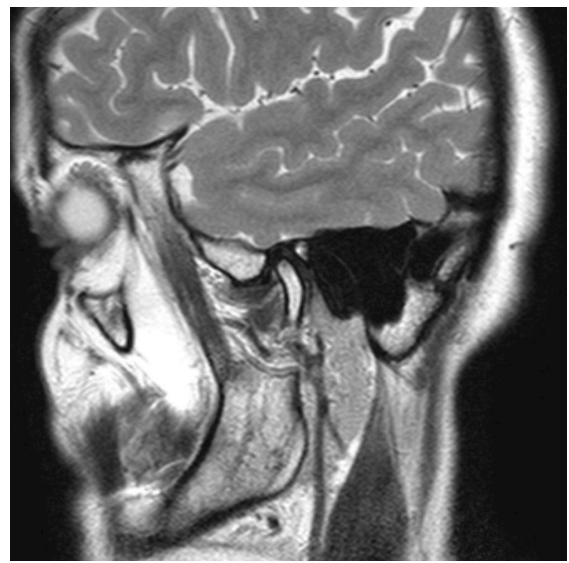

Figure 3 – An example of input MRI-image

Dataset annotation was performed on the Roboflow platform [19]. Despite the availability of the automatic annotation function, its use turned out to be unsuitable due to the low contrast of the structures of the maxillofacial area. As a result, manual segmentation was performed with the allocation of two classes of objects: "TMJ" and "JAW".

The limited sample size (94 images) required the use of augmentation methods to increase data diversification. Using the Roboflow tools, all images were resized to 640x640 pixels, the contrast was adjusted, and the following transformations were applied:

- brightness adjustment (± 20%);
- exposure change (± 15%).

As a result of augmentation, the dataset volume was increased to 232 images. Annotations for training the models were presented in the format of image contour coordinates. An example of an annotated image is shown in Figure 4.

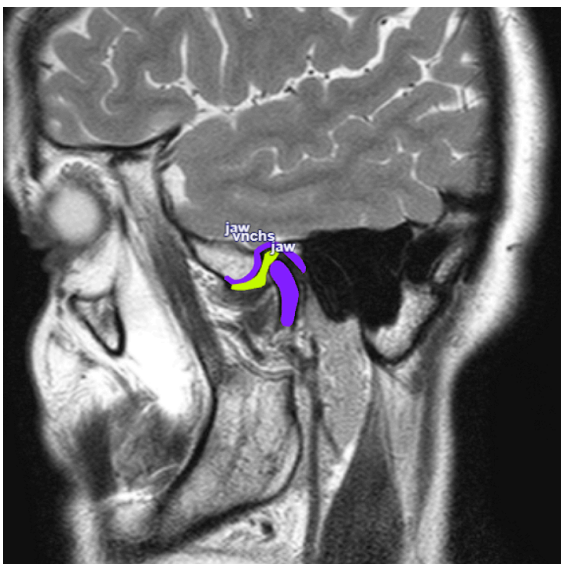

Figure 4 – An example of annotated image

## 2.2 Metrics for a comparative analysis

To conduct a comparative analysis of the effectiveness of different architectures of neural networks, we selected standardized metrics that are often used in such studies [18, 20]:

- Dice Score is a metric that shows how well the predicted mask matches the true mask, and it is calculated using the formula
  $F1 = 2*TP / (2*TP + FP + FN)$,
  where TP is the number of true positive examples, FN is the number of false negative examples, FP is the number of false positive examples.
- Mean Average Precision (mAP) is the average value of Average Precision (AP) for all detected classes. It is calculated as the Average Precision for all classes and then averaged.
  $mAP = 1/n * sum(AP)$,
  where n is the number of classes
- Precision – the proportion of correctly predicted pixels relative to all predicted pixels;
- Sensitivity – the proportion of correctly predicted pixels relative to all true pixels.
  $Sensitivity = TP / (TP + FN)$,
  where TP is the number of correctly

predicted pixels, FP is the number of false negative pixels

- Specificity is a metric that shows how well the model differentiates the background from the object.

$Specificity = TN / (TN + FP)$, where TN is the number of pixels correctly attributed to the background, and FP is the number of pixels erroneously predicted as an object.

## 3. Results

The U-Net with the pre-trained EfficientNet encoder was chosen as the basic architecture for solving the segmentation problem. The model was trained on an NVIDIA graphics processing unit (GPU) for 50 epochs, and the total training time was 55 minutes. The validation metrics are presented below:

- Dice Score: 0.7320
- Precision: 0.7710
- Sensitivity : 0.7225
- Specificity: 0.9971
- mAP: 0.6593

The result of the U-Net model is presented in Figure 5. The trained model was able to indicate the lower and upper jaws of the articular disc itself, but the boundaries were not clearly defined.

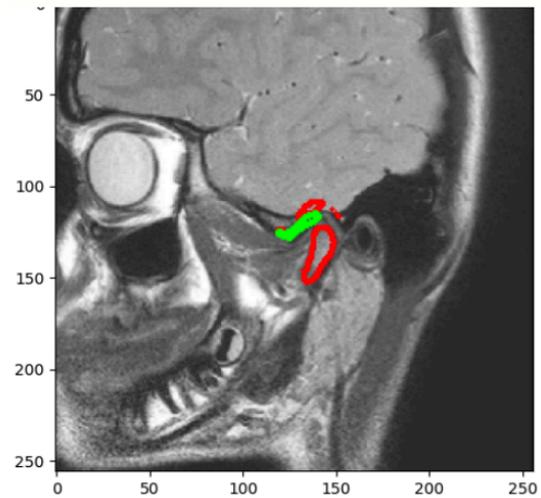

Figure 5 – The U-net model segmentation result

The next model was YOLO8n-seg, which was trained in the Google Colab environment for 100 epochs and took only 7 minutes. The segmentation results are presented in Figure 6. The validation metrics are as follows:

- Dice Score: 0.7189
- Precision: 0.7740
- Sensitivity: 0.6711
- Specificity: 0.9822
- mAP: 0.6798

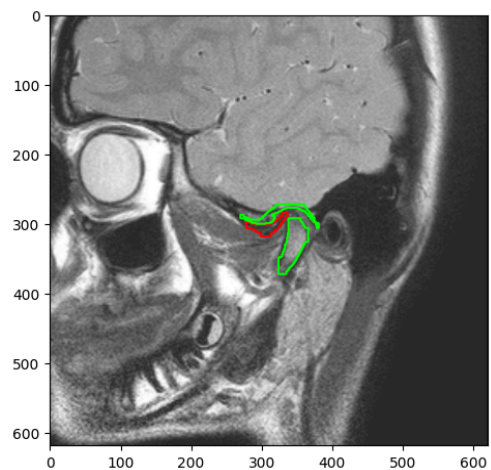

Figure 6 – The YOLO8n-seg segmentation result

After the YOLO8n-seg model, a new model from the YOLO family was used – YOLO11l-seg [21]. The model was also trained in Google Colab for 100 epochs, the training took 24 minutes. The metrics are almost the same as for YOLO8n-seg:

- Dice Score: 0.7158
- Precision: 0.7658
- Sensitivity: 0.6719
- Specificity: 0.9837
- mAP: 0.7250

Figure 7 shows the clear boundaries of the lower and upper jaw and the articular disc itself.

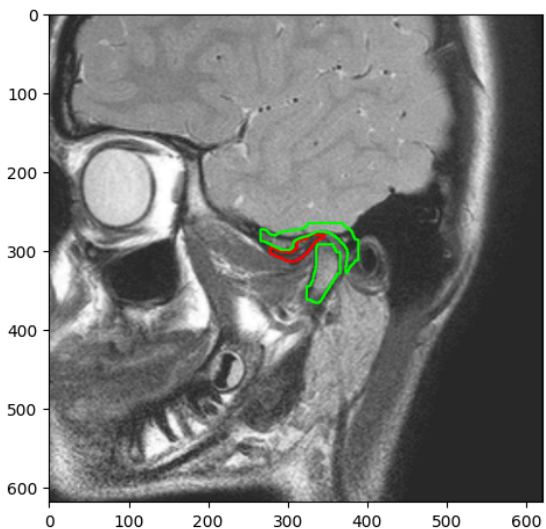

Figure 7 - The YOLO11l-seg segmentation result

The last model, as an experiment, was the model proposed by the Roboflow service itself, with which we marked up MRI scans. The model was trained for 30 minutes, during 89 epochs. In general, we got good results:

- Dice Score: 0.85
- Precision: 0.906
- Sensitivity: 0.785
- Specificity: 0.98
- mAP: 0.901

The model accurately recognized the articular disc and the boundaries of the jaw (Figure 8).

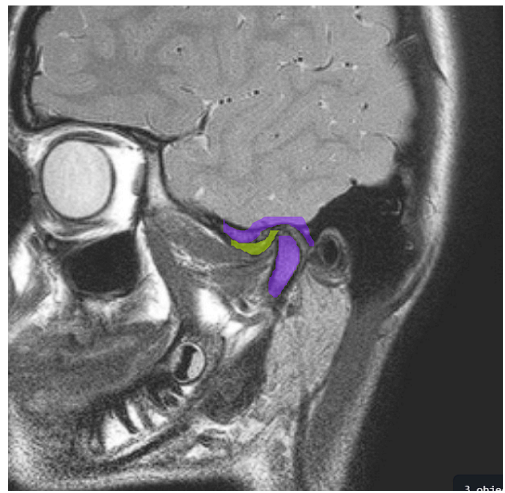

Figure 8 – The Robloflow segmentation result

### 4. Discussion

Table 1 presents a comparative analysis of the models used in the study according to the key metrics: Dice Score, Precision, Sensitivity, Specificity, mAP. Some of these metrics for evaluating models were used by Korean scientists in their study on the topic "Multiclass segmentation of the temporomandibular joint using collective deep learning" [18]. However, they used completely different models. We decided to use different models for training and to check how effectively they recognize TMJ on MRI images in comparison with existing approaches.

Table 1 – Comparison of the results of the considered models

|  | Dice Score | Precision | Sensitivity | Specificity | mAP |
|---|---|---|---|---|---|
| U-Net | 0.73 | 0.77 | 0.72 | 0.99 | 0.66 |
| YOLO8n-seg | 0.71 | 0.77 | 0.67 | 0.98 | 0.68 |
| YOLO11l-seg | 0.72 | 0.77 | 0.67 | 0.98 | 0.73 |
| Roboflow 3.0 Instance Segmentation (Fast) | 0.85 | 0.9 | 0.79 | 0.98 | 0.9 |

According to the analysis, the results show that the Roboflow model has demonstrated high efficiency. Compared to other models, Roboflow showed the best results according to the derived metrics and was able to accurately determine the boundaries of the TMJ disc and jaw on the test images (Figure 9). It can be assumed that this architecture may well be recommended by us for the segmentation of similar medical images.

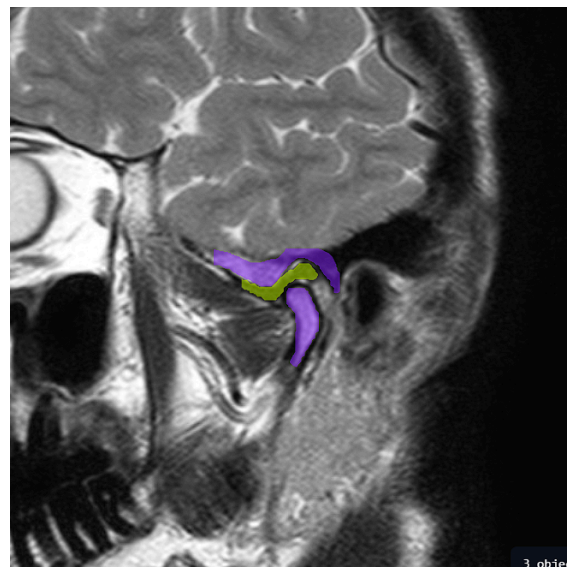

Figure 9 – Detection of the TMJ-disk and jaw on a test image using Roboflow 3.0 Instance Segmentation

**5. Conclusion**

During the study, the existing solutions for the study of the temporomandibular joint were analyzed, but it was found that they are not suitable for studying its key element - the articular disc, the damage or displacement of which can lead to unpleasant consequences for a person.

To develop our own solution, we collected and labeled a dataset with MRI scans, trained several models and finally identified the model that best copes with the task - the recognition of the temporomandibular joint disc. This model is Roboflow 3.0 Instance Segmentation (Fast), as it showed the best results among other models according to the selected metrics during the study. It was able to clearly define the boundaries of both the TMJ and the boundaries of the upper and lower jaws.

Based on the results obtained, we developed a scheme for identifying TMJ pathologies (Figure 10), which includes the following stages: MRI image segmentation, calculation of the distance between the upper and lower jaw (using the algorithm), drawing conclusions from the results obtained and identifying the pathology, as well as displaying this information on the doctor-user's screen.

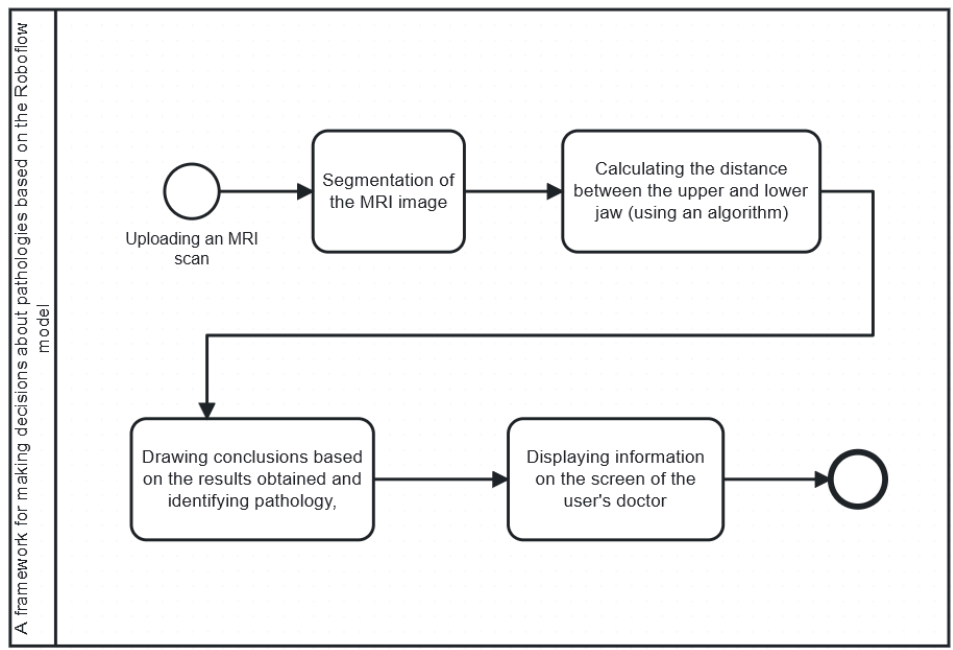

Figure 10 – Framework for deriving solutions to pathologies based on the Roboflow model

The prospect of further research in the context of the above framework for the determination of pathologies will be the development of an algorithm for measuring the distance between the upper and lower jaw, as well as determining the position of the articular disc in the joint to identify the pathologies of the TMJ.

This study was supported by the Ministry of Science and Higher Education of the Russian Federation within the framework of a State assignment (FEWZ-2024-0052).